\documentstyle[11pt]{article}

\def \lket {|}
\def \rket {\rangle}
\def \lbra {\langle}
\def \rbra {|}
\def\rank {rank}
\def\ufour{{\overline{4}}}
\newcommand{\ket}[1]{\lket #1\rket}
\newcommand{\bra}[1]{\lbra #1\rbra}
\newtheorem{Definition}{Definition}
\newtheorem{Theorem}{Theorem}
\newtheorem{Lemma}{Lemma}
\newtheorem{Claim}{Claim}
\newcommand{\comment}[1]{}

\newcommand{\proof}{\noindent {\bf Proof: }}
\newcommand{\qed}{\nobreak \ifvmode \relax \else
      \ifdim\lastskip<1.5em \hskip-\lastskip
      \hskip1.5em plus0em minus0.5em \fi \nobreak
      \vrule height0.75em width0.5em depth0.25em\fi}

\def\bbbc{{\bf C}}
\def\bbbr{{\bf R}}

\def\>{\rangle}
\def\<{\langle}

\def\bfact{\begin{fact}}
\def\efact{\end{fact}}

\def\bv{\left( \begin{matrix}}
\def\ev{\end{matrix} \right)}

\def\be{\begin{equation}}
\def\ee{\end{equation}}
\def\bes{\begin{eqnarray}}
\def\ees{\end{eqnarray}}
\def\bess{\begin{eqnarray*}}
\def\eess{\end{eqnarray*}}

\begin{document}

\title{Quantum $t$-designs: 
$t$-wise independence in the quantum world}

\author{Andris Ambainis\thanks{
Department of Combinatorics and Optimization and the
Institute for Quantum Computing, University of Waterloo,
{\tt ambainis@uwaterloo.ca}. Supported by NSERC, CIAR, ARO, MITACS and IQC
University Professorship.}
and 
Joseph Emerson\thanks{Department of Applied Mathematics and the Institute for
Quantum Computing, University of Waterloo, {\tt jemerson@uwaterloo.ca}.
Supported by NSERC and MITACS.}}

\date{}
\maketitle

\begin{abstract} 
A $t$-design for quantum states is a finite set of quantum states
with the property of simulating the Haar-measure on quantum states
w.r.t. any test that uses at most $t$ copies of a state. 
We give efficient constructions for approximate 
quantum $t$-designs for arbitrary $t$.
We then show that an approximate 4-design provides a derandomization
of the state-distinction problem considered by Sen (quant-ph/0512085), 
which is relevant to solving certain instances of
the hidden subgroup problem.
\end{abstract}

\section{Introduction}

$t$-wise independent and approximately $t$-wise independent probability
distributions have been extremely useful in combinatorics and the theory
of computing. In this paper, we study their quantum counterparts, 
{\em quantum $t$-designs}.

Intuitively, a quantum $t$-design is a probability distribution over
quantum states which cannot be distinguished from the uniform probability
distribution over all quantum states (the Haar measure) if we are given
$t$ copies of a state from this probability distribution.
More formally, we define

\begin{Definition}
\label{def:exact-new} [Generalization of the definition in
Ref.~\cite{Renes}] A probability distribution over quantum states
$(p_i, \ket{\phi_i})$ is a complex projective $(t, t)$-design if
\[ \sum_i p_i (\ket{\phi_i}\bra{\phi_i})^{\otimes t}=
\int_{\psi} (\ket{\psi}\bra{\psi})^{\otimes t} d \psi ,\] where the
integral over $\ket{\psi}$ on the left hand side is taken over the
Haar measure on the unit sphere in $\bbbc^N$.
\end{Definition}

This definition of complex-projective $(t,t)$-designs, or
\emph{quantum t-designs} has been previously studied in two contexts.

In the context of quantum information theory,
\cite{Barnum,Renes,KR05,KRSW05,Dankert} have studied quantum
2-designs, giving constructions of 2-designs with $O(N^2)$ states
and applying them to various problems in quantum information.
Hayashi et al. \cite{HHH} gave a construction
of a $t$-design for arbitrary $t$ with $O(t)^N$ of states.
This is efficient for a fixed dimension $N$ but inefficient when $N$ 
is much larger than $t$. The $N=2$ (one-qubit) case was independently
solved by Iblisdir and Roland \cite{IR}.

Second, quantum $t$-designs are related to 
$t$-designs of vectors on the unit sphere in
$\bbbr^N$, called \emph{spherical t-designs}, which
have been studied in the mathematics literature since a seminal
paper by Delsarte, Goethals and Seidel \cite{DGS}. An inefficient 
construction of
an exact spherical $t$-design with $t^{O(N^2)}$ vectors has been
given by Bajnok \cite{Bajnok} and Korevaar and Meyers \cite{KM}. 
A spherical $t$-design in $\bbbr^N$ can be
transformed into a $(t/2, t/2)$-design in $\bbbc^{N/2}$. Thus, those
results also imply the existence of quantum $t$-designs with a
similar number of states.

To summarize the previous work (for the case when $t$ is fixed and 
the dimension $N$ is large), inefficient constructions of 
quantum $t$-designs with an exponential
number of states are known for any $t$ and efficient constructions
are known for $t=2$. The contributions of this paper are as
follows:
\begin{enumerate}
\item
We introduce the notion of an approximate $t$-design;
\item
We give an efficient construction of approximate $t$-designs, 
with $O(N^{3t})$ states,
for any $t$;
\item
We show how to apply an approximate 4-design to derandomize the 
state-distinction result by Sen \cite{Sen}.
\end{enumerate}

\section{Summary of results}

Our definition of an approximate
$(t, t)$-design is as follows:

\begin{Definition}
\label{def:approx-new}
A probability distribution over quantum states
$(p_i, \ket{\phi_i})$ is an $\epsilon$-approximate $(t, t)$-design
if
\[ (1-\epsilon) \int_{\psi} (\ket{\psi}\bra{\psi})^{\otimes t} d \psi \leq
\sum_i p_i (\ket{\phi_i}\bra{\phi_i})^{\otimes t} \leq
(1+\epsilon) \int_{\psi} (\ket{\psi}\bra{\psi})^{\otimes t} d \psi ,\]
where the integral over $\ket{\psi}$ on the left hand side
is taken over the Haar measure on the unit sphere in $\bbbc^N$
and
\begin{equation}
\label{eq:POVM}
\sum_i p_i \ket{\phi_i}\bra{\phi_i} =
\int_{\psi} \ket{\psi}\bra{\psi} d \psi .
\end{equation}
\end{Definition}

Instead of requiring closeness to a $t$-design in the $l_{\infty}$
norm, as in Definition \ref{def:approx-new}, one could use a
different norm (e.g., $l_1$ or $l_2$-norm). This might make design
easier to construct but closeness in $l_1$ or $l_2$ is not
sufficient for Theorem \ref{thm:design} and, possibly, other applications.

\begin{Theorem}
\label{thm:twise}
Fix a constant $t$. Then, for every $N\geq 2t$, there exists an
$O(\frac{1}{N^{1/3}})$-approximate
$(t, t)$-design consisting of $O(N^{3t})$ quantum states\footnote{The
big-O constants can depend on $t$.}.
\end{Theorem}

The $t$-design of Theorem \ref{thm:twise} 
can be efficiently implemented, for several
meanings of ``efficiently implemented'':
\begin{enumerate}
\item
It is possible to generate a quantum state $\ket{\phi}$ distributed according 
to the probability distribution
$(p_i, \ket{\phi_i})$ in time $O(\log^c N)$.
\item
Because of equation (\ref{eq:POVM}), the operators 
$N p_i \ket{\phi_i}\bra{\phi_i}$ form a POVM measurement.
This POVM measurement can be implemented in time $O(\log^c N)$.
\end{enumerate}
The first property is just the normal definition of being able to sample
from the probability distribution. (Since we are dealing with
states in $N$ dimensions, which can be described by $\log N$ qubits,
``efficient'' means polynomial in $\log N$.) The second
property may seem unusual at first but it is exactly what we need
for our application (Theorem \ref{thm:design}).

In section \ref{sec:improved}, 
we show that the number of states in the $\epsilon$-approximate
$(t, t)$-design can be decreased to
$O(N^t \log^c N)$.
There is a simple way to generate the states in
the resulting $(t, t)$-design but we are not sure if the 
corresponding POVM measurement can be efficiently implemented.

We now give the application of Theorem \ref{thm:twise}.
Radhakrishnan et al. \cite{RRS} have shown

\begin{Theorem}
\label{thm:sen} Let $\ket{\psi_1}, \ket{\psi_2}$ be two orthogonal
quantum states in $\bbbc^N$. Then,
\[ E_{\hat M} \| \hat M(\psi_1) - \hat M(\psi_2) \|_1 = \Omega(1) ,\]
where $\hat M$ is an orthonormal basis picked uniformly at random
from the Haar measure.
\end{Theorem}

This result was improved by Sen\cite{Sen}.

\begin{Theorem}
\label{thm:sen1} Let $\rho_1, \rho_2$ be two mixed states in
$\bbbc^N$ with $\rank \rho_1+\rank \rho_2 \leq \frac{\sqrt{N}}{K}$
for a sufficiently large $K$. Then,
\[ E_{\hat M} \| \hat M(\rho_1) - \hat M(\rho_2) \|_1 = \Omega(f) ,\]
where $\hat M$ is an orthonormal basis picked uniformly at random
from the Haar measure and $f=\|\rho_1-\rho_2\|_F$ is the Frobenius
norm of $\rho_1-\rho_2$ ($\|A\|_F=\sqrt{\sum_{k, l=1}^N
|a_{kl}|^2}$).
\end{Theorem}

Theorem \ref{thm:sen} is a particular case of Theorem
\ref{thm:sen1}, since
$\|\ket{\psi_1}\bra{\psi_1}-\ket{\psi_2}\bra{\psi_2}\|_F = 2$.
As next theorem shows, we can replace the measurement in 
a random orthonormal basis by a POVM w.r.t. a complex projective
4-design (where ``POVM with respect to $(p_i, \ket{\phi_i})$'' is just 
the POVM consisting of one-dimensional projectors 
$N p_i \ket{\phi_i}\bra{\phi_i}$).

\begin{Theorem}
\label{thm:design} Let $f=\|\rho_1-\rho_2\|_F$ and $\epsilon< c
f^4$, where $c$ is a sufficiently small constant. Then, for any
mixed states $\rho_1$, $\rho_2$ in $\bbbc^{N \times N}$,
\[ \| \hat M(\rho_1) - \hat M(\rho_2) \|_1 = \Omega(f) ,\]
where $\hat M$ is a POVM with respect to an $\epsilon$-approximate
$(4, 4)$-design.
\end{Theorem}

Theorems \ref{thm:design} and \ref{thm:twise} together
derandomize Theorem \ref{thm:sen1}, as long as $f=\Omega(n^{-1/12})$.

The rest of this paper is structured as follows.
In section \ref{sec:defs}, we compare different definitions of $t$-designs.
Then, in section \ref{sec:construct}, we give our construction
of approximate $t$-designs (Theorem \ref{thm:twise}). 
In section \ref{sec:derandomize}, we show how to use
(4, 4)-designs for state-distinction (Theorem \ref{thm:design}).
These are the main results of our paper.

Some more 
technical claims are postponed to appendices. In appendix \ref{app:haar},
we derive expressions for expected values of monomials involving
amplitudes of a quantum state drawn from Haar measure. 
In appendix \ref{app:def}, we prove two theorems relating definitions
of $(t, t)$-designs and in appendix \ref{sec:efficient},
we show how to implement POVM w.r.t. our $(t, t)$-design efficiently.

\section{Definitions of $(t, t)$-designs}
\label{sec:defs}

The earlier papers on $(t, t)$-designs used a different definition
of $(t, t)$-designs, in terms of polynomials of the amplitudes of
a state $\ket{\phi_i}$. In this section, we show that the two definitions
are equivalent. We also present a condition on the polynomials of
amplitudes which implies Definition \ref{def:approx-new}.

Let $p(x_1, \ldots, x_N, y_1, \ldots, y_N)$ be a polynomial of degree at most
$t$ in variables $x_1, \ldots, x_N$ and degree at most $t$ in variables
$y_1, \ldots, Y_N$. For a state $\ket{\psi}=\sum_{j=1}^N \alpha_j \ket{j}$,
we define
\[ p(\psi)=p(\alpha_1, \ldots, \alpha_N, \alpha^*_1, \ldots, \alpha_N^*) .\]

\begin{Definition}
\label{def:exact-old}
A probability distribution over quantum states
$(p_i, \ket{\phi_i})$ is a complex projective $(t, t)$-design
if, for arbitrary polynomial $p(x_1, \ldots, x_N, y_1, \ldots, y_N)$
of degree $t$ in variables $x_1, \ldots, x_N$ and degree $t$
in variables $y_1, \ldots, y_N$, we have
\begin{equation}
\label{eq:design} \int_{\psi} p(\psi)d \psi =\sum_i p_i p(\phi_i),
\end{equation}
where the integral over $\ket{\psi}$ on the left hand side
is taken over the Haar measure on the unit sphere in $\bbbc^N$.
\end{Definition}

\begin{Theorem}
\label{thm:equiv}
$(p_i, \ket{\phi_i})$ is a complex projective $(t, t)$-design
according to Definition \ref{def:exact-new} if and only if it is
a complex projective $(t, t)$-design according to 
Definition \ref{def:exact-old}.
\end{Theorem}

\proof
In appendix \ref{app:def}.
\qed

As shown in the Appendix \ref{app:haar}, 
the Haar-expectation of any unbalanced
monomial $p$ (i.e. monomial with $d_i\neq c_i$ for some $i\in\{1,
\ldots, N\}$) is
\begin{equation}
\label{eq:mono}
\int_{\psi} p(\psi)d \psi = 0
\end{equation}
and the Haar-expectation of any monomial of the form
\[ p=\prod_{j=1}^k x_{i_j}^{c_j} (x^*_{i_j})^{c_j} \]
for distinct $x_{i_1}, \ldots, x_{i_k}$ is
\begin{equation}
\label{eq:mono1}
\int_{\psi} p(\psi)d \psi = \frac{c_1!\ldots c_k!}{N(N+1)\ldots(N+d-1)}
\end{equation}

We show that having an approximate version of these requirements is
sufficient for an approximate $(t, t)$-design:
\begin{Theorem}
\label{thm:approx-imply}
Assume that a probability distribution over quantum states
$(p_i, \ket{\phi_i})$ satisfies the following constraints:
\begin{enumerate}
\item
$\sum_i p_i p(\phi_i)=0$ for any unbalanced monomial $p$,
\item
\[ \left| \sum_i p_i p(\phi_i) - \frac{c_1!\ldots c_k!}{N(N+1)\ldots(N+d-1)}
\right|
\leq \epsilon \frac{c_1!\ldots c_k!}{N(N+1)\ldots(N+d-1)} .\]
for monomials $p=\prod_{j=1}^k x_{i_j}^{c_j} (x^*_{i_j})^{c_j}$ with
$c_1+\ldots+c_k=d$ and $d\leq t$ and
\item
$\sum_i p_i p(\phi_i)=\frac{1}{N}$ for monomials
$p=x_j x^*_j$, where $j\in\{1, 2, \ldots, N\}$.
\end{enumerate}
Then, $(p_i, \ket{\phi_i})$
is an $t! \epsilon$-approximate $(t, t)$-design.
\end{Theorem}

\proof
In appendix \ref{app:def}.
\qed

\section{Constructing approximate $(t, t)$-designs}
\label{sec:construct}

\subsection{Main construction}
\label{sec:main-approx}

In this section, we prove Theorem \ref{thm:twise}.
It suffices to construct a set of states that satisfies the requirements
of Theorem \ref{thm:approx-imply}.
For simplicity, assume that $N$ is a power of 2.
We use

\begin{Theorem}
\cite{Zuckerman} For any $N=2^k$, there is a set $S$ of $N^{d}$
functions $f:\{0, \ldots, N-1\}\rightarrow\{0, \ldots, N-1\}$ such
that, for any distinct $k_1, \ldots, k_d\in\{0, \ldots, N-1\}$, the
probability distribution of $f(k_1), \ldots, f(k_d)$ (where $f$ is
chosen uniformly at random from $S$) is exactly the uniform
distribution over tuples of $d$ elements of $\{0,\dots,N-1\}$.
\end{Theorem}

Such $S$ are called {\em d-wise} independent sets.
The second technical tool that we use is the Gaussian quadrature.

\begin{Lemma}
\label{lem:Gauss}
\cite[Chapter 10.6]{Hamming}
Let $X$ be a real-valued random variable and
$c_i=E[X^i]$, for $i\in\{1, \ldots, 2t\}$.
Then, there exist real $q_1, \ldots, q_{2t}$ and $x_1, \ldots, x_{2t}$
such that $q_i\geq 0$, $q_1+\ldots+q_{2t}=1$ and
\[ \sum_{i=1}^{2t} q_i x_i^j = c_j ,\]
for all $j\in\{1, \ldots, 2t\}$.
\end{Lemma}

In other words, for any continuous probability distribution, we can
always construct a discrete probability distribution with the first $2t$
moments having the same values.

Let $\ket{\psi}=\sum_{i=1}^N \alpha_i \ket{i}$ be a state drawn from
the Haar measure. Let $P_N$ be the probability distribution of
$\alpha_1$ and let $P=\lim_{N\rightarrow\infty} \sqrt{N} P_N$.
Let $\alpha$ be drawn from $P$.
We let $X=|\alpha|$ with probability 1/2 and
$X=-|\alpha|$ with probability 1/2. Then, $E[X^j]=0$ for odd $j$
and
\[ E[X^j]=\lim_{N\rightarrow \infty} N^{j/2} \cdot\frac{(j/2)!}{N(N+1)
\ldots (N+j/2-1)} = (j/2)! \]
for even $j$.
We apply Lemma \ref{lem:Gauss} to get $q_1, \ldots, q_{2t}$
and $x_1, \ldots, x_{2t}$.

We then replace each $q_i$ with one of the two closest multiples
of $1/N$ (i.e., $\frac{\lfloor N q_i \rfloor}{N}$ or
$\frac{\lfloor N q_i \rfloor+1}{N}$) so that
$q_1+\ldots+q_{2t}$ remains 1.
We simultaneously adjust $x_i$ so that $q_i x_i^2$ stays the same.
This changes the probabilities $q_j$ by at most $1/N$
and $x_j$ by at most a factor of $1+O(1/N)$.
The moments $\sum_i q_i x_i^j$ change by at most
\[ \frac{2t}{N} (\max_i x^j_i-\min_i x^j_i) = O\left( \frac{1}{N} \right) \]
due to change in probabilities $x_i$ and
at most a multiplicative factor of $(1+O(1/N))^t=1+O(1/N)$
due to change in $x_j$.
Thus, we have

\begin{Claim}
\label{claim:rounding}
\[ \left| E[X^{2c_j}] - c_j!\right| =
O\left(\frac{1}{N}\right) .\]
\end{Claim}

Let $S_1$ be a set of $t$-wise independent functions
$f:\{1, \ldots, N\}\rightarrow\{1, \ldots, N\}$
and $S_2$ be a set of $2t$-wise independent functions
$g:\{1, \ldots, N\}\rightarrow\{1, \ldots, N\}$.

We now consider the set of quantum states $\ket{\psi_{f, g}}=
\sum_{j=1}^N \alpha_{f, g, j} \ket{j}$,
(where $f\in S_1$, $g\in S_2$) generated in a following way:
\begin{enumerate}
\item
Let $\beta_{f,g,j}$ be a complex number with absolute value
$a_{f, j}=\frac{x_l}{\sqrt{N}}$ where
$l$ is such that $q_1+\ldots+q_{l-1}< \frac{f(j)}{N} \leq
q_1+\ldots+q_{l}$
and amplitude $e^{i \pi g(j)/N}$.
\item
Let
\[ \alpha_{f,g,j} = \frac{\beta_{f,g,j}}{\sqrt{\sum_{i=1}^N a_{f, i}^2}} ,\]
for $j\in\{1, \ldots, N\}$.
\item
Let
\[ p_{f, g}=\frac{\sum_{i=1}^N a_{f, i}^2}{|S_1|\cdot |S_2|} .\]
\end{enumerate}

We claim that $(p_{f, g}, \ket{\psi_{f, g}})$ is an approximate $(t,
t)$-design. We first show

\begin{Claim}
\label{claim:fix-f}
Fix $f\in S_1$.
If we pick $\ket{\psi}=\sum_{j=1}^N \alpha_{f,g,j} \ket{j}$
uniformly at random
from $\ket{\psi_{f, g}}$, $g\in S_2$, then
\begin{enumerate}
\item
$E[h]=0$ for any unbalanced monomial $h$ of degree at most $2t$;
\item
\begin{equation}
\label{eq:unbalanced}
E[h]= \prod_{j=1}^k \left( \frac{a_{f, i_j}}{\sqrt{\sum_{i=1}^N
a_{f, i}^2}}
\right)^{2 c_j}
\end{equation}
for a balanced monomial
\[ h=(\alpha_{i_1}\alpha^*_{i_1})^{c_1}
\ldots (\alpha_{i_k}\alpha^*_{i_k})^{c_k} \]
of degree at most $2t$.
\end{enumerate}
\end{Claim}

\proof
Let
\[ h=\alpha^{c_1}_{i_1}(\alpha^*_{i_1})^{d_1}
\ldots \alpha^{c_k}_{i_k}(\alpha^*_{i_k})^{d_k}\]
be a monomial of degree $d=\sum_j (c_j+d_j)$ which is at most $2t$. Then,
it contains at most
$2t$ different variables $\alpha_{i_1}$, $\ldots$, $\alpha_{i_k}$.
Since $g$ is picked from a $2t$-wise independent family of functions,
phase of every of those variables is an independent random variable
$y_j$ taking values $1, e^{i\pi/N}$, $\ldots$, $e^{i(2N-1)\pi/N}$ and
\[ E[h] = \prod_{j=1}^k \left( \frac{a_{f, i_j}}{\sqrt{\sum_{i=1}^N
a_{f, i}^2}}
\right)^{c_j+d_j} \prod_{j=1}^k E[e^{i\pi y_j(c_j-d_j)/N}] .\]
If, for some $j$, $c_j\neq d_j$, then the corresponding expectation
$E[e^{i\pi y_j(c_j-d_j)/N}]$ is 0. This proves the first part of the
claim.
Otherwise, all expectations are 1 and the second part follows.
\qed

The first part of the claim immediately implies that the first requirement of
Theorem \ref{thm:approx-imply} is satisfied.
To prove the second requirement, we first observe two facts about
the normalization factor $\sqrt{\sum_{i=1}^N a_{f, i}^2}$:
\begin{enumerate}
\item
We have
\[ N \min_j x^2_j \leq \sum_{i=1}^N a_{f, i}^2 \leq N \max_j x^2_j,\]
because each of $a_{f, i}$ is equal to one of $x_j$.
\item
We have
\begin{equation}
\label{eq:Cheb}
Pr_f \left[ \left|\sum_{i=1}^N a_{f, i}^2  - 1 \right| \geq \frac{C}{\sqrt{N}}
\right] \leq \frac{1}{C^2}.
\end{equation}
To prove this bound, we first observe that
$D[a_{f, i}^2]=E[a_{f, i}^4]-E^2[a_{f, i}^2]=2!-(1!)^2=1$.
Since the variables $a_{f, i}^2$ are $t$-wise independent (and, hence, 2-wise independent),
we have $D[\sum_{i=1}^N a_{f, i}^2]=N$. The bound now follows from Chebyshev
inequality.
\end{enumerate}
We have to bound the expectation of the random variable
\begin{equation}
\label{eq:xprime}
X'= \prod_{j=1}^k \left( \frac{a_{f, i_j}}{\sqrt{\sum_{i=1}^N a_{f, i}^2}}
\right)^{2c_j} ,
\end{equation}
with $a_{f, i}=f(i)$, where $f\in S_1$ and the probability of $f$ is
equal to $|S_2|p_{f, g}=\frac{\sum_{i=1}^N a_{f, i}^2}{|S_1|}$.
Equivalently, we can bound the expectation of
\begin{equation}
\label{eq:x}
X =\frac{\prod_{j=1}^k a_{f, i_j}^{2c_j}}{(\sum_{i=1}^N a_{f, i}^2)^{d-1}} ,
\end{equation}
when each $f$ is picked with probability $\frac{1}{|S_1|}$.
We observe that
\[ \frac{1}{N^d} \frac{\min^{2d}_j x_j}{\max^{2d-2}_j x_j} < X
< \frac{1}{N^d} \frac{\max^{2d}_j x_j}{\min^{2d-2}_j x_j}  .\]
Thus, the maximum and the minimum value
of $X$ differ by at most $\frac{D}{N^{d}}$ where
$D= \frac{\min^{2d}_j x_j}{\max^{2d-2}_j x_j}-
\frac{\max^{2d}_j x_j}{\min^{2d-2}_j x_j}$ is independent of $N$.
We take $C=N^{1/6}$. If $|\sum_{i=1}^N a_{f, i}^2  - 1 | \leq
\frac{C}{\sqrt{N}}$,
then
\[ \left( 1-\frac{1}{N^{1/3}} \right)^{d-1} \prod_{j=1}^k a_{f, i_j}^{2c_j}
\leq X \leq
\left( 1+\frac{1}{N^{1/3}} \right)^{d-1} \prod_{j=1}^k a_{f, i_j}^{2c_j} .\]
Therefore,
\[ \left( 1-\frac{1}{N^{1/3}} \right)^{d-1}
E\left[ \prod_{j=1}^k a_{f, i_j}^{2c_j} \right]
- \frac{D}{N^{d}} Pr\left[\left|\sum_{i=1}^N a_{f, i}^2  - 1 \right| \leq
\frac{C}{\sqrt{N}}\right]
\leq E[X] \]
\[ \leq
\left( 1+\frac{1}{N^{1/3}} \right)^{d-1} E\left[\prod_{j=1}^k
a_{f, i_j}^{2c_j}\right] +
\frac{D}{N^{d}} Pr\left[\left|\sum_{i=1}^N a_{f, i}^2  - 1 \right| \leq
\frac{C}{\sqrt{N}}\right]  .\]
By equation (\ref{eq:Cheb}),
\[
Pr\left[\left|\sum_{i=1}^N a_{f, i}^2  - 1 \right| \leq \frac{C}{\sqrt{N}}
\right] \leq \frac{1}{C^2}.
\]
Together with the independence of random variables $a_{f, i_j}^{2c_j}$ (which is implied by
$t$-wise independence of $a_{f, i_j}$ and $k\leq t$), this implies
\[ \left( 1-\frac{1}{N^{1/3}} \right)^{d-1} \prod_{j=1}^k
E[a_{f, i_j}^{2c_j}] - \frac{D}{N^{d+1/3}}
\leq E[X]\leq \left( 1+\frac{1}{N^{1/3}} \right)^{d-1} \prod_{j=1}^k
E[a_{f, i_j}^{2c_j}] +
\frac{D}{N^{d+1/3}}. \]
The theorem now follows from claim \ref{claim:rounding}.

To prove third requirement of Theorem \ref{thm:approx-imply},
let $\ket{\psi_{f,g}}=\sum_{i=1}^N \alpha_{f,g,i} \ket{i}$.
Then, the expectation of $x_j x^*_j$ is
\[ \sum_{f,g} p_{f,g} \alpha_{f, g, i} \alpha^*_{f, g, i} =
\sum_{f,g} \frac{\sum_{i=1}^N |\beta_{f,g,i}|^2}{|S_1|\cdot|S_2|}
\frac{|\beta_{f,g,i}|^2}{\sum_{i=1}^N |\beta_{f,g,i}|^2} =
\sum_{f,g} \frac{|\beta_{f,g,i}|^2}{|S_1|\cdot|S_2|} \]
which is just the expectation of $|\beta_{f,g,i}|^2$ when $f,g$ are chosen
uniformly at random.
This expectation is $\sum_{l=1}^{2t} q_l \frac{x^2_l}{N}
= \frac{1}{N}$, by the definition of
the random variables $x_j$.
This completes the proof of Theorem \ref{thm:twise}.

\subsection{Improved construction}
\label{sec:improved}

To decrease the number of states in the $(t, t)$-design, we
use a result about approximately $t$-wise families of functions.

\begin{Definition}
A family of functions $f:\{0, \ldots, N-1\}\rightarrow\{0, \ldots, m-1\}$ is
$t$-wise $\delta$-dependent if, for any pairwise
distinct $i_1, \ldots, i_t\in\{0, \ldots, N-1\}$,
the variational distance between the probability distribution
of $f(i_1), \ldots, f(i_t)$ and the uniform distribution on $\{0, \ldots, m-1\}^t$
is at most $\delta$.
\end{Definition}

\begin{Theorem}
\label{thm:naor}
\cite{Naor}
There is a family of $t$-wise $\delta$-dependent functions
$f:\{0, \ldots, N-1\}\rightarrow\{0, 1\}$
of cardinality $2^{O(t+\log \log N+\log \frac{1}{\delta})}$.
\end{Theorem}

Instead of 0-1 valued functions, we will need $m$-valued functions.
If $m=2^k$, we can use the construction of \cite{Naor} to construct
a $t \log m$-wise $\delta$-dependent family of functions
$f':\{0, \ldots, N \log m-1 \}\rightarrow \{0, 1\}$
of cardinality $2^{O(t \log m +\log \log N+ \log \frac{1}{\delta})}$.
We then define $f(i)$ to be equal to the number formed by
$f'(i \log N)$, $f'(i\log N+1)$, $\ldots$, $f'(i\log N+\log N-1)$.
This gives us a $t$-wise $\delta$-dependent family of functions
$f:\{0, \ldots, N -1 \}\rightarrow \{0, \ldots, m-1\}$.

We also use

\begin{Theorem}
\label{thm:alon}
\cite{ABI}
There is a family of $t$-wise independent functions
$f:\{0, \ldots, N-1\}\rightarrow\{0, 1\}$
of cardinality $O(N^{t/2})$.
\end{Theorem}

We modify the previous construction in a following way:
\begin{itemize}
\item
$f$ is picked from a $t$-wise $\delta$-dependent
family of functions $f:\{0, \ldots, N-1\}\rightarrow\{0, \ldots, m-1\}$
(where $m$ will be specified later).
By the discussion after Theorem \ref{thm:naor},
such a family has cardinality $O(m^{ct} (\delta \log N)^c)$
for some constant $c$.
\item
$g$ is replaced by two functions: $g_1:\{0, \ldots, N-1\}\rightarrow \{0, 1\}$
and $g_2:\{0, \ldots, N-1\}\rightarrow \{0, \ldots, N-1\}$.
$g_1$ is picked from a $2t$-wise independent family of functions
of Theorem \ref{thm:alon}.
$g_2$ is picked from a $t$-wise $\delta$-dependent family
of functions $\{0, \ldots, N-1\}\rightarrow \{0, \ldots, m-1\}$.
Instead of $e^{2 i \pi g(j)/N}$, our phase is
$(-1)^{g_1(j)} e^{2 i \pi g_2(j)/m}$.
\end{itemize}
The number of states in our sample size is then $O(N^t m^{ct} ( \log N/\delta)^c)$
for some constant $c$.
We will take $\delta=O(\epsilon)$ and $m=\Omega(1/\epsilon)$.
This gives a design with $O(N^t (\log N/\epsilon)^c)$ states.
We claim that this gives us an $\epsilon$-approximate $(t, t)$-design.

The argument is the same as in section \ref{sec:main-approx},
with the following changes:
\begin{enumerate}
\item
In Claim \ref{claim:rounding}, we have $O(\frac{1}{m N^{c_j}})$ instead
of $O(\frac{1}{N^{c_j+1}})$.
\item
In Claim \ref{claim:fix-f},
for monomials that contain $\alpha^c_i (\alpha^*_i)^d$ with $c+d$ odd,
the expectation is still exactly 0, because of the $(-1)^{g_1(j)}$
multiplier which is 1 with probability 1/2 and -1 with probability 1/2.
For monomials
\begin{equation}
\label{eq:even}
h=\alpha^{c_1}_{i_1}(\alpha^*_{i_1})^{d_1}
\ldots \alpha^{c_k}_{i_k}(\alpha^*_{i_k})^{d_k}
\end{equation}
with $c_1+d_1$, $\ldots$, $c_k+d_k$ all even, the
$(-1)^{g_1(j)}$ term is always 1.
Since $g_2$ is $t$-wise $\delta$-dependent,
the expectation of $h$ for a fixed $f$ deviates from
the expectation for $t$-wise independent $g_2$ by at most
$c\delta$ times
\begin{equation}
\label{eq:even-expect}
\prod_{j=1}^k \left( \frac{a_{f, i_j}}{\sqrt{\sum_{i=1}^N
a_{f, i}^2}} \right)^{c_j+d_j}
\end{equation}
for some constant $c$.
\item
We then have to bound the expectation of $X'$ (equation (\ref{eq:xprime}))
which can be replaced by the expectation of $X$ (equation (\ref{eq:x})) in
the same way as before.
The only change is that $f$ is now chosen from a $t$-wise $\delta$-dependent
distribution. This means that the expectation of
\[ \prod_{j=1}^k \left( \frac{a_{f, i_j}}{\sqrt{\sum_{i=1}^N
a_{f, i}^2}} \right)^{c_j+d_j} \]
differs from the expectation under a $t$-wise independent
distribution by at most $\frac{D \delta}{N^d}$.
\end{enumerate}

Overall, this introduces an additional error of
order $c(\max \frac{1}{m}, \delta)$ times the
expectation of (\ref{eq:even-expect}). Thus, choosing
$\delta=O(\epsilon)$ and $m=\Omega(1/\epsilon)$
with appropriate constants is sufficient for
an $\epsilon$-approximate design.

To complete the proof, we have to verify that Theorem 
\ref{thm:approx-imply} works, when, instead of an assumption about 
balanced terms, we just have an assumption about terms of the form 
(\ref{eq:even}). This part is mostly technical and is omitted 
in this version.

\section{Derandomizing the measurement in a random basis}
\label{sec:derandomize}

In this section, we prove theorem \ref{thm:design}.
First, we consider the case when we have an exact $(4, 4)$-design instead of
an approximate one.
By the definition of POVM,
\[  \|  \hat M(\rho_1) - \hat M(\rho_2) \|_1  =
 \sum_{j=1}^L N p_j \left|
\bra{\phi_j} \rho_1 -\rho_2 \ket{\phi_j} \right|  \]
where $L$ is the number of states in the $(4, 4)$-design.
Theorem \ref{thm:design} now follows from

\begin{Lemma}
\label{lem:design}
\[ \sum_{j=1}^L p_j  \left|
\bra{\phi_j} \rho_1 -\rho_2 \ket{\phi_j}  \right|
= \Omega\left(\frac{f}{N}\right) .\]
\end{Lemma}

To prove Lemma \ref{lem:design}, we use the fourth moment method
of Berger \cite{Berger}:

\begin{Lemma}
\cite{Berger}
\label{lem:berger}
For any random variable $S$,
\[ E[|S|]\geq \frac{E[S^2]^{3/2}}{E[S^4]^{1/2}} .\]
\end{Lemma}

This means that
\[ \sum_{j=1}^L p_j  \left|
\bra{\phi_j} \rho_1 -\rho_2 \ket{\phi_j}  \right|
\frac{(\sum_{j=1}^L p_j  \left|
\bra{\phi_j} \rho_1 -\rho_2 \ket{\phi_j}  \right|^2)^{3/2}}{
(\sum_{j=1}^L p_j  \left|
\bra{\phi_j} \rho_1 -\rho_2 \ket{\phi_j}  \right|^4)^{1/2}} .\]
We now bound the numerator and the
denominator of this expression.
We first observe that
\begin{equation}
\label{eq:deg2} \sum_{j=1}^L p_j | \bra{\phi_j} \rho_1 -\rho_2
\ket{\phi_j} |^2 = E_{\ket{\phi}} | \bra{\phi} \rho_1 -\rho_2
\ket{\phi} |^2
\end{equation}
with $\ket{\phi}$ on the right hand side chosen according to the Haar measure.
(Let $\ket{\phi}=\sum_{i=1}^N \alpha_i \ket{i}$. Then, equation (\ref{eq:deg2})
is true because $| \bra{\phi} \rho_1 -\rho_2 \ket{\phi} |^2$
is a polynomial of degree 2 in variables $\alpha_i$ and degree 2 in variables
$\alpha^*_i$ and, therefore, its expectation is the same for Haar measure and
for a $(4, 4)$-design.)
Let $\lambda_1, \ldots, \lambda_n$ be the
eigenvalues of $\rho_1-\rho_2$. Then,
\begin{equation}
\label{eq:0sum}
\lambda_1+\ldots+ \lambda_N = Tr (\rho_1)- Tr (\rho_2)=0,
\end{equation}
\[ \lambda^2_1+\ldots+\lambda^2_N  = \|\rho_1-\rho_2\|_F = f^2 .\]
For the moment, assume that $\rho_1-\rho_2$ is diagonal in the basis
$\ket{1}$, $\ldots$, $\ket{N}$ and $\ket{i}$ is the eigenvector with
the eigenvalue $\lambda_i$. By writing out $g(\phi)=( \bra{\phi}
\rho_1 -\rho_2 \ket{\phi} )^2$ for $\ket{\phi}=\sum_{i=1}^N \alpha_i
\ket{i}$, we get that $g(\phi)$ is equal to
\[ \sum_{i=1}^N \lambda^2_i (\alpha_i \alpha_i^*)^2 +
2\mathop{\sum_{i,j=1}^N}_{i<j}
\lambda_i \lambda_{i-1} \alpha_i \alpha^*_i \alpha_j \alpha^*_j \]
plus some unbalanced terms.
When $\ket{\phi}$ is picked from the Haar measure, the expectation of
each unbalanced term is 0. The expectation of balanced terms
is given by equation (\ref{eq:mono1}). This gives us
\[ E_{\ket{\phi}}[g]= \frac{2}{N(N+1)} \sum_{i=1}^N \lambda^2_i +
\frac{1}{N(N+1)} 2\mathop{\sum_{i,j=1}^N}_{i<j}
 \lambda_i \lambda_{i-1} = \]
\begin{equation}
\label{eq:exf}
\frac{1}{N(N+1)} \left( \sum_{i=1}^N \lambda_i \right)^2 +
\frac{1}{N(N+1)} \sum_{i=1}^N \lambda^2_i =
\frac{f^2}{N(N+1)} .
\end{equation}
If $\rho_1-\rho_2$ is not diagonal in the
basis $\ket{1}, \ldots, \ket{N}$, let $U$ be a unitary transformation that
maps $\ket{1}, \ldots, \ket{N}$ to the eigenbasis of $\rho_1-\rho_2$.
Then,
\[ E ( \bra{\phi} \rho_1 -\rho_2 \ket{\phi} )^2 =
E (\bra{\phi} U^{\dagger} (\rho_1 -\rho_2) U \ket{\phi} )^2 ,\]
by the invariance of Haar measure under unitary transformations
and $U^{\dagger} (\rho_1 -\rho_2) U$ is diagonal in the basis
$\ket{1}, \ldots, \ket{N}$.
Thus, the expression (\ref{eq:exf}) for the expectation remains
the same even if $\rho_1-\rho_2$ is not diagonal in the
basis $\ket{1}, \ldots, \ket{N}$.

The expectation of $g$ must be the same if $\ket{\phi}$ is picked
from a $(4, 4)$-design. Therefore,
\begin{equation}
\label{eq:deg2new}
\sum_{j=1}^L p_j (\lbra \phi_j |\rho_1 - \rho_2|\phi_j \rket )^2 =
\frac{f^2}{N(N+1)} .
\end{equation}
Similarly,
\begin{equation}
\label{eq:deg4new} \sum_{j=1}^L p_j ( \lbra \phi_j |\rho_1- \rho_2|
\phi_j \rket  )^4 = E (\lbra \phi |\rho_1 - \rho_2| \phi \rket )^4 =
E \left( \sum_{i=1}^n \lambda_i \alpha_i \alpha^*_i \right)^4 .
\end{equation}
For the second equality, we again assumed that $\rho_1-\rho_2$ is diagonal in the
basis $\ket{1}, \ldots, \ket{N}$. This assumption can be removed in the same
way as before.

Denote $v_i=\lambda_i \alpha_i \alpha_i^*$. Let $N^\ufour$ be a
shortcut for $N(N+1)(N+2)(N+3)$. Then, (\ref{eq:deg4new}) is equal
to
\[ E \left[ \sum_{i=1}^N v^4_i +
\mathop{\sum_{i,j=1}^N}_{i\neq j} 4 v^3_i v_j +
\mathop{\sum_{i,j=1}^N}_{i < j} 6 v^2_i v^2_j +
\mathop{\sum_{i,j,k=1}^N}_{j < k, i\neq j, i\neq k} 12 v^2_i v_j v_k +
\mathop{\sum_{i,j,k,l=1}^N}_{i<j<k<l} 24 v_i v_j v_k v_l \right] = \]
\[ \frac{24}{N^\ufour} \left( \sum_{i=1}^n \lambda^4_i +
\mathop{\sum_{i,j=1}^N}_{i\neq j} \lambda^3_i \lambda_j +
\mathop{\sum_{i,j=1}^N}_{i < j} \lambda^2_i \lambda^2_j +
\mathop{\sum_{i,j,k=1}^N}_{j < k, i\neq j, i\neq k} \lambda^2_i \lambda_j
\lambda_k + \mathop{\sum_{i,j,k,l=1}^N}_{i<j<k<l}
\lambda_i \lambda_j \lambda_k \lambda_l \right) = \]
\[ \frac{1}{N^\ufour} \left( \sum_{i=1}^N \lambda_i \right)^4 +
\frac{6}{N^\ufour} \left( \sum_{i=1}^N \lambda_i \right)^2
\left( \sum_{i=1}^N \lambda^2_i \right) + \]
\[ \frac{8}{N^\ufour} \left( \sum_{i=1}^N \lambda_i \right)
\left( \sum_{i=1}^N \lambda^3_i \right) +
\frac{9}{N^\ufour} \sum_{i=1}^N \lambda^4_i +
\frac{6}{N^\ufour} \mathop{\sum_{i,j=1}^N}_{i < j}
\lambda^2_i \lambda^2_j = \]
\begin{equation}
\label{eq:deg4}
\frac{9}{N^\ufour} \sum_{i=1}^N \lambda^4_i +
\frac{6}{N^\ufour} \mathop{\sum_{i,j=1}^N}_{i < j}
\lambda^2_i \lambda^2_j \leq
\frac{9}{N^\ufour}  \left( \sum_{i=1}^N \lambda^2_i \right)^2 =
\frac{9}{N^\ufour} f^4,
\end{equation}
with the first equality following from equation (\ref{eq:mono1}),
the second equality following by a rearrangement of terms, the
third equality following from (\ref{eq:0sum}) and the inequality following
by expanding $\left( \sum_{i=1}^N \lambda^2_i \right)^2$ and using
$\lambda_i^2 \lambda_j^2\ge 0$ for all $i, j$.

By combining Lemma \ref{lem:berger} and equations (\ref{eq:deg2new}),
(\ref{eq:deg4new}) and (\ref{eq:deg4}), we get
\[ \sum_{j=1}^L p_j  \left|
\bra{\phi_j} \rho_1 -\rho_2 \ket{\phi_j}  \right| \geq
\frac{(\frac{f^2}{N(N+1)})^{3/2}}{(\frac{9f^4}{N(N+1)(N+2)(N+3)})^{1/2}}
\geq \frac{f}{3N} .\]
\comment{
Without the loss of generality,
we can assume that the vectors $\ket{\phi_j}$ are
ordered in the order of decreasing
$\lbra \phi_j |\rho_1 -\rho_2| \phi_j \rket$.
Let $j_0$ be the maximum number for which
\[ (\lbra \phi_{j_0} |\rho_1 - \rho_2| \phi_{j_0} \rket)^2
\geq \frac{18 f^2}{(N+2)(N+3)}. \]
Let
\[ S=\sum_{j=1}^{j_0} p_j
(\lbra \phi_{j} |\rho_1 -  \rho_2| \phi_j \rket )^2
.\]
Then,
\[ \sum_{j=1}^{j_0} p_j
(\lbra \phi_{j} |\rho_1- \rho_2|\phi_j \rket  )^4
\geq  \]
\[ \frac{18 f^2}{(N+2)(N+3)} \sum_{j=1}^{j_0} p_j
(\lbra \phi_{j} |\rho_1 - \rho_2|\phi_j \rket )^2 =
\frac{18 f^2}{(N+2)(N+3)} S .\]
Together with equation (\ref{eq:deg4}), this implies
$S\leq \frac{f^2}{2N(N+1)}$
and, together with equation (\ref{eq:deg2}), that means
\[ \sum_{j=j_0+1}^L p_j
(\lbra \phi_{j} |\rho_1-\rho_2|\phi_j \rket)^2
\geq
\frac{f^2}{2N(N+1)}  \]
and
\[ \sum_{j=j_0+1}^L p_j
\left| \lbra \phi_{j} |\rho_1 - \rho_2|\phi_j \rket \right|
 \geq
\frac{\sum_{j=j_0+1}^L p_j
(\lbra \phi_{j} |\rho_1 - \rho_2|\phi_j \rket)^2}{
|\lbra \phi_{j_0+1} |\rho_1- \rho_2|\phi_{j_0+1} \rket |} \geq \]
\[ \frac{\frac{f^2}{2N(N+1)}}{\sqrt{\frac{18 f^2}{(N+2)(N+3)}}}
\geq \frac{f}{\sqrt{72} N} .\]
}
This implies Lemma \ref{lem:design} in the case when we have an exact
$(4, 4)$-design.

For the case of $\epsilon$-approximate $(4, 4)$-design,
we bound the difference between the expectations
of $(\lbra \phi | \rho_1-\rho_2 |\phi \rket)^2$
and $(\lbra \phi | \rho_1-\rho_2 |\phi \rket)^4$
when $\ket{\phi}$ is drawn from the Haar measure and when
it is drawn from an approximate $(4, 4)$-design.
We can rewrite
\begin{equation}
\label{eq:eigensum} (\lbra \phi | \rho_1-\rho_2 |\phi \rket)^2 =
\sum_{j} \lambda_j (\lbra \phi | \varphi_j \rket \lbra \varphi_j |\phi \rket)^2
\end{equation}
where $\lambda_j$ are eigenvalues of $\rho_1-\rho_2$ and
$\ket{\varphi_j}$ are the corresponding eigenvectors.
By the definition of $(t, t)$-design,
the expectation of
\[ (\lbra \phi | \varphi_j \rket \lbra \varphi_j |\phi \rket)^2 =
\bra{\varphi_j^{\otimes 2}} (\ket{\phi}\bra{\phi})^{\otimes 2}
\ket{\varphi_j^{\otimes 2}}   \]
changes by at most $\frac{\epsilon}{M}$ (where $M$ is the dimension
of symmetric subspace ${\cal H}_{sym}$ for 2 copies of a state $\ket{\phi}$)
when $\ket{\phi}$ is picked from an $\epsilon$-approximate $(4, 4)$ design.
The entire sum (\ref{eq:eigensum}) changes by at most
\[ \sum_j |\lambda_j | \frac{\epsilon}{M} \leq \frac{2\epsilon}{M} ,\]
where the inequality follows from the sum of all positive eigenvalues
of $\rho_1-\rho_2$ being at most $Tr \rho_1 =1$ and the sum of
absolute values of negative eigenvalues being at most $Tr \rho_2=1$.
Since $M\geq \frac{N^2}{2}$, this is at most $\frac{4\epsilon}{N^2}$.

Similarly, we can show that the expectation of $(\lbra \phi | \rho_1-\rho_2 |\phi \rket)^4$
changes by at most $\frac{2\cdot 4!\epsilon}{N^4}$ when
$\ket{\phi}$ is picked from an $\epsilon$-approximate $(4, 4)$ design.
For our proof to work,
those changes have to be small compared to the expectations of
$(\lbra \phi | \rho_1-\rho_2 |\phi \rket)^2$
and $(\lbra \phi | \rho_1-\rho_2 |\phi \rket)^4$
when $\ket{\phi}$ is picked from the Haar measure.
This happens if $\epsilon< c f^4$ for a sufficiently
small constant $c$.

\subsection{$(2, 2)$-designs are not sufficient}

We note that using a $(4, 4)$-design is essential for
our construction. First, as shown by Berger \cite{Berger},
a bound on the fourth moment is necessary to obtain a bound
on $E[|S|]$. Second, some well-known $(2, 2)$-designs
are insufficient for distinguishing between some orthogonal
quantum states.

For example, this is true for $(2, 2)$-designs constructed
from mutually unbiased bases \cite{KR05}.
Let $\ket{\phi_1}$, $\ldots$, $\ket{\phi_N}$ be an
orthonormal basis for an $N$-dimensional Hilbert space and
$\ket{\varphi_1}$, $\ldots$, $\ket{\varphi_N}$ be another
orthonormal basis for the same space.
The two bases are {\em mutually unbiased} if
$|\lbra\phi_i |\varphi_j \rket|=\frac{1}{\sqrt{N}}$ for all $i, j$.
If $N$ is prime, there exist $N+1$ orthonormal bases
$\ket{\phi_{i, 1}}$, $\ldots$, $\ket{\phi_{i, N}}$
(for $i\in\{1, \ldots, N+1\}$)
such that any two of them are mutually unbiased.
The collection of states $\ket{\phi_{i, j}}$
(with probabilities $1/N(N+1)$ each) is then a $(2, 2)$-design
\cite{KR05}.

We now consider the POVM corresponding to this $(2, 2)$-design.
This POVM is equivalent to randomly choosing $i\in\{1, \ldots, N+1\}$
(with probabilities $1/(N+1)$ each)
and then performing an orthogonal measurement in the basis
$\ket{\phi_{i, 1}}$, $\ldots$, $\ket{\phi_{i, N}}$.

Let $\ket{\psi_1}=\ket{\phi_{1, 1}}$, $\ket{\psi_2}=\ket{\phi_{1, 2}}$.
Then, measuring in the first basis perfectly distinguishes the
states $\ket{\psi_1}$ and $\ket{\psi_2}$ but measuring
either of those states in any other basis $\ket{\phi_{i, 1}}$, $\ldots$,
$\ket{\phi_{i, N}}$ produces the uniform probability distribution.
Therefore, performing the POVM on $\ket{\psi_1}$ and
$\ket{\psi_2}$ produces two probability distributions with
the variational distance $2/(N+1)$ between them.

However, since $\ket{\psi_1}$ and $\ket{\psi_2}$ are orthogonal,
we have $\|\ket{\psi_1}\bra{\psi_1}-\ket{\psi_2}\bra{\psi_2}\|_F=2$.

\section{Open problems}

It appears plausible that the methods developed above can be applied
to construct approximate $t$-designs for unitary transformations
(defined in \cite{Dankert}). An important set of open questions is
whether the efficient approximate t-designs developed above can be
applied to derandomize other protocols that make use of random
states and/or random unitary operators, for example, the protocol
for locking classical correlations \cite{HL+}.

{\bf Acknowledgments.}
We thank Oded Regev for suggesting to use Gaussian quadrature
and Robin Blume-Kohout, Aram Harrow, Debbie Leung, Pranab Sen and Andreas 
Winter for discussions and comments on this paper.

\newpage
\begin{appendix}

\noindent
{\Huge Appendix}

\section{Haar Average of State-Component Monomials}
\label{app:haar}

Consider the $N$-dimensional Hilbert space, $\mathcal{H}= \bf{C}^N$
consisting of the set of normalized pure quantum states. These
states correspond to the points of a unit sphere $S^{2N-1}$ which is
the ``surface'' of a ball in $2N$ real dimensions.  If we remove the
arbitrary and unphysical phase associated with each state then we
are left with the complex projective space $CP^{N-1}$. In either
case there exists a unique 'natural' measure that is induced by the
invariant (Haar) measure on the unitary group $U(N)$: a uniformly
random pure state can be defined by the action of a uniformly random
unitary matrix on an arbitrary reference state, $|\phi\> =
U|\phi_0\>$. The measure on pure states is distinguished by the
rotational invariance of the Haar measure. This measure, which I
will denote $\mu(\psi)$, is equivalent to the uniform measure on the
unit sphere $S^{2N-1}$. Choosing a fixed representation, $| \psi \>
= \sum_i c_i | i \rangle$, the uniform measure for normalized
vectors (pure states) in $\mathcal{H}$ can be expressed using the
Euclidean parametrization,
\begin{equation}\label{Euclideanpsi}
 d\mu(\psi \in S^{2N-1}) = \left(\Pi_{i=1}^N d^2 c_i\right)
 \delta\left(\sum_{l=1}^N |c_l|^2 -1\right)
 \end{equation}
where $\delta$ is the Dirac delta function.

The average value of any function $f: \mathcal{H} \to \bf{C}$ takes
the explicit form,
\begin{eqnarray}\label{ave}
    \langle f(\psi) \rangle_\psi & =  \frac{1} { V_{S^{2N-1}}} \int_{S^{2N-1}}  \; f(\psi)
    \; d\mu(\psi \in S^{2N-1}).
\end{eqnarray}
Often the function can be represented or approximated as a
polynomial in the components of the pure state. The terms of such a
polynomial may be calculated directly using the Euclidean measure
(\ref{Euclideanpsi}) by using an integration trick \cite{Ullah}.
Consider first calculating the volume of pure states. We have,
\begin{eqnarray}
 V_{S^{2N-1}} & = & \int_{\psi \in S^{2N-1}} d\mu(\psi) \nonumber \\
 & = & \int_{\psi \in S^{2N-1}}
 \left(\Pi_{i=1}^N d^2 u_i\right) r^{-2N+1}
 \delta\left(  \sqrt{\sum |u_l|^2} -r \right)
\end{eqnarray}
where we have made the change of variables $c_i = u_i/ r$, and used
the identity $\delta(a/b-1) = b \delta(a-b)$. (Notice that for
calculating the volume of the $2N-1$ sphere one has to be careful
about distinguishing the constraint $\delta\left(\sum_{l=1}^N
|u_l|^2 -r\right)$ from $\delta(\sqrt{ \sum_{l=1}^N |u_l|^2} -r) $
for the variable radius  $r^2 = \sum_{l=1}^N |c_l|^2$.) Collecting
factors of $r$ on the left hand side, the main trick for evaluating
this integral is to introduce the integrating factor $\exp(-r^2)dr$
and then integrate both sides with respect to $r$,
\bes
 V_{S^{2N-1}} \int_0^{\infty} dr \; r^{2N-1} \exp(-r^2) & = & \int
 \Pi_{i=1}^N d^2 u_i \; e^{-\sum_{j=1}^N |u_j|^2} \nonumber \\
 V_{S^{2N-1}} \frac{\Gamma(N)}{2} & = & \left[\Gamma\left(\frac{1}{2}\right)\right]^{2N} = \pi^N \nonumber \\
V_{S^{2N-1}}  & = & \frac{2 \pi^N}{(N-1)!}
\ees
where we've used the integral identity $\int_0^\infty r^q e^{-r^2} =
(1/2)\Gamma(q+1/2)$, and recovered the well-known result for the
volume of the unit $R$ sphere in $R+1$ real dimensions: $V_{S^{R}} =
2 \pi^{R/2}/(R/2-1)!$ with $R = 2N$.

Now we can calculate the correlation function for a $k$-body product
of distinct state components,
\be
I(k,t) \equiv \< |c_1 |^{2t_1} |c_2 |^{2t_2} \cdots |c_k |^{2t_k} \>
= \frac{1}{V_{S^{2N-1}}} \int_{\psi \in S^{2N-1}} d\mu(\psi)\; |c_1
|^{2t_1} |c_2 |^{2t_2} \cdots |c_k |^{2t_k},
\ee
which corresponds to the expectation of a homogeneous polynomial of
degree $(t,t)$, where $t= \sum_{j=1}^k t_j$. By the same method as
above we obtain,
\bes
I(k,t) \; V_{S^{2N-1}}\; \int_0^{\infty} dr \; r^{2N-1+2\sum t_j}
\exp(-r^2) & = & \int
 \Pi_{i=1}^N d^2 u_i \; e^{-\sum_{l=1}^N |u_l|^2} \Pi_{j=1}^k  |u_j|^{2t_j}\nonumber \\
I(k,t) \frac{2 \pi^N}{(N-1)!} \frac{\Gamma(N+\sum t_j)}{2}
& = & \left[ \int d^2 u \; e^{-|u|^2} \right]^{N-k} \; \Pi_{j=1}^k \int d^2
u_j \; e^{-|u_j|^2} |u_j|^{2t_j} \nonumber \\
I(k,t)
& = & \frac{(N-1)!}{\pi^{k}(N+t-1)!} \; \Pi_{j=1}^k \int d^2
u_j \; e^{-|u_j|^2} |u_j|^{2t_j}.
\ees
where in the last line we've used
\be
\left[ \int d^2 u \; e^{-|u|^2}  \right]^{N-k} = \left( \pi
\right)^{N-k}.
\ee
In order to evaluate the remaining factor we change to polar
coordinates, with $u_j = x+iy$, and $dxdy = r dr d\theta$, giving
for each $u_j$ the factor,
\be
\int d^2 u_j \; e^{-|u_j|^2} |u_j|^{2t_j} = 2\pi \int_0^\infty dr \;
e^{-|u_j|^2} r^{2t_j} = 2\pi \Gamma(t_j +1)/2 = \pi t_j!.
\ee
Hence,
\be
\< |c_1 |^{2t_1} |c_2 |^{2t_2} \cdots |c_k |^{2t_k} \> = \frac{t_1!
t_2! \cdots t_k!}{(N+t-1)(N+t-2) \cdots (N)}.
\ee

\section{Proofs of Theorems from section 3}
\label{app:def}

\proof [of Theorem \ref{thm:equiv}]

In Definition \ref{def:exact-old}, instead of a general polynomial $p$,
it suffices to consider the case when $p$ is a monomial
\[ x_1^{c_1} x_2^{c_2} \ldots x_N^{c_N} y_1^{d_1}
y_2^{d_2} \ldots y_N^{d_N}. \]
If the equation (\ref{eq:design}) is true for all monomials $p$,
it will also be true for all polynomials $p$.

To see the equivalence with Definition \ref{def:exact-new},
observe that each entry of the density matrix
$\sum_i p_i (\ket{\phi_i}\bra{\phi_i})^{\otimes t}$ is
an expectation of a monomial in the amplitudes of $\ket{\phi_i}$
and the corresponding entry of
$\int_{\psi} (\ket{\psi}\bra{\psi})^{\otimes t} d \psi$ is
the expectation of the same monomial when $\ket{\psi}$ is picked
from the Haar measure. Thus, if the expectations are the same
for any monomial, the density matrices
$\sum_i p_i (\ket{\phi_i}\bra{\phi_i})^{\otimes t}$
and $\int_{\psi} (\ket{\psi}\bra{\psi})^{\otimes t} d \psi$
are the same and Definition \ref{def:exact-new} holds.

In the other direction, for every monomial of degree $t$
in variables $\alpha_i$ and degree $t$ in variables $\alpha^*_i$, there is
an entry in $\int_{\psi} (\ket{\psi}\bra{\psi})^{\otimes t} d \psi$
which is equal to its expectation. Therefore,
Definition \ref{def:exact-new} also implies Definition \ref{def:exact-old}.
\qed

\proof [of Theorem \ref{thm:approx-imply}]
Let ${\cal H}_{sym}$ be the subspace spanned by all states of the form
$\ket{\psi}^{\otimes t}$.

Then, $\int_{\psi} (\ket{\psi}\bra{\psi})^{\otimes t} d \psi$
is just $\frac{I}{M}$, the completely mixed state over the
subspace ${\cal H}_{sym}$ (where $M=\dim {\cal H}_{sym}$).
We need to prove that, for any $\ket{\psi_{sym}}\in {\cal H}_{sym}$,
\begin{equation}
\label{eq:ip}
\frac{1- \epsilon}{M} \leq E \left[ \bra{\psi_{sym}}
 \left( p_i \ket{\phi_i} \bra{\phi_i}^{\otimes t} \right)
\ket{\psi_{sym}} \right] \leq \frac{1+\epsilon}{M} .
\end{equation}
We can write any state $\ket{\psi_{sym}}\in {\cal H}_{sym}$ as
\[ \ket{\psi_{sym}} = \sum_{i_1 \leq i_2 \leq \ldots \leq i_t}
\alpha_{i_1, \ldots, i_t} \ket{\psi_{i_1, \ldots, i_t}} ,\]
where $\ket{\psi_{i_1, \ldots, i_t}}$ is the uniform superposition
over all basis states $\ket{j_1, \ldots, j_t}$ such that
the multisets $\{i_1, \ldots, i_t\}$ and $\{j_1, \ldots, j_t\}$ are
equal.

Let $d_{i_1, \ldots, i_t}$ be the number of different
basis states $\ket{j_1, \ldots, j_t}$ such that
the multiset $\{j_1, \ldots, j_t\}$ is equal to $\{i_1, \ldots, i_t\}$.
(If there are $k$ different elements in $\{i_1, \ldots, i_t\}$,
occurring $c_1, \ldots, c_k$ times, then
$d_{i_1, \ldots, i_t}= \frac{t!}{c_1! \ldots c_k !}$.)
Then, each of $\ket{j_1, \ldots, j_t}$ has the amplitude
of $\frac{1}{\sqrt{d_{i_1, \ldots, i_t}}}$ in the state
$\ket{\psi_{i_1, \ldots, i_t}}$.
Let $\alpha_1, \ldots, \alpha_N$ be the amplitudes of a
state $\ket{\psi}=\sum_j \alpha_j \ket{j}$ which is picked from the
distribution $(p_i, \ket{\phi_i})$.
Then,
\[ \lbra \psi_{i_1, \ldots, i_t} \ket{\psi_{i}} =
d_{i_1, \ldots, i_t} \cdot \frac{1}{\sqrt{d_{i_1, \ldots, i_t}}}
\lbra i_1, \ldots, i_t \ket{\phi_i} =
\sqrt{d_{i_1, \ldots, i_t}} \alpha_{i_1} \ldots \alpha_{i_t} .\]
By summing over all components $\ket{\psi_{i_1, \ldots, i_t}}$ of
$\ket{\psi_{sym}}$, we get that
\[ \bra{\psi_{sym}}
 \ket{\phi_i} \bra{\phi_i}^{\otimes t}
\ket{\psi_{sym}} = \]
\begin{equation}
\label{eq:ip-sum}
\left( \sum_{i_1\leq \ldots \leq i_t} \alpha^*_{i_1, \ldots, i_t}
\sqrt{d_{i_1, \ldots, i_t}} \alpha_{i_1} \ldots \alpha_{i_t} \right)
\left( \sum_{j_1 \leq \ldots \leq j_t} \alpha_{j_1, \ldots, j_t}
\sqrt{d_{j_1, \ldots, j_t}} \alpha^*_{j_1} \ldots \alpha^*_{j_t} \right)
.
\end{equation}

If $\ket{\psi}$ was picked from the Haar measure,
the expectation of (\ref{eq:ip-sum}) would be $\frac{1}{M}$.
Thus, it suffices to bound the difference of
the expected value of (\ref{eq:ip-sum}) between the two cases:
$\ket{\psi}$ picked from Haar measure and
$\ket{\psi}$ picked from $(p_i, \ket{\phi_i})$.

We expand both sums in the equation (\ref{eq:ip-sum}).
If $j_l\neq i_l$ for some $l\in\{1, \ldots, t\}$,
then the expectation of a term
\begin{equation}
\label{eq:term}
\alpha_{i_1} \ldots \alpha_{i_t} \alpha^*_{j_1} \ldots \alpha^*_{j_t}
\end{equation}
is 0 in both cases. When $j_l = i_l$ for all $l\in\{1, \ldots, k\}$,
the expectations of (\ref{eq:term}) under both distributions differ by
at most
\[ \epsilon \frac{c_1!\ldots c_k!}{N(N+1)\ldots(N+d-1)}
d_{i_1, \ldots, i_t}
|\alpha_{i_1, \ldots, i_t}|^2 .\]
Since the squared amplitudes $|\alpha_{i_1, \ldots, i_t}|^2$ sum up to 1,
this means that the difference between expectation of (\ref{eq:ip-sum})
in the two cases is at most
\[ \epsilon \frac{c_1!\ldots c_k!}{N(N+1)\ldots(N+d-1)}
d_{i_1, \ldots, i_t} =
\epsilon \frac{t!}{N(N+1) \ldots (N+d-1)} .\]
\qed

\section{Efficient implementation}
\label{sec:efficient}

In this section, we show how to implement (an approximation) of the
POVM w.r.t. one-dimensional projectors
$E_{f,g}=p_{f, g} N \ket{\psi_{f, g}} \bra{\psi_{f, g}}$
efficiently. 

For this construction, we will need to use the particular
family of $t$-wise independent functions from \cite{Zuckerman}.
Let $G$ be a finite field with $N$ elements (which exists because
we constrained $N$ to be power of 2).
We associate $\{0, \ldots, N-1\}$ with the elements of $G$.
Using the construction of \cite{Zuckerman} results in
$f(x)$ ranging over all polynomials (in $x$) over $G$
of degree at most $t-1$ and $g(x)$ ranging over all polynomials over
$G$ of degree at most $2t-1$.

We have $E_{f, g}= E_g E_f$ where
\begin{enumerate}
\item
$E_f$ is a diagonal matrix,
with the entries $(E_f)_{j, j} = \frac{1}{N^{t-1}}
a_{f, j}^2$
on the diagonal;
\item
$E_g=\frac{1}{N^{2t-1}}\ket{\psi_g}\bra{\psi_g}$
where $\ket{\psi_g}= \frac{1}{\sqrt{N}}
\sum_{l=0}^{N-1} e^{2\pi i \frac{g(j)}{N} }\ket{j}$.
\end{enumerate}
Both $E_f$ and $E_g$ constitute a POVM.
Thus, we can perform the measurement in two steps,
first measuring $E_f$ and then measuring $E_g$.

\comment{
Instead of measuring $E_f$, we implement the POVM consisting of $E'_f$
where
\[ E'_f = \frac{1}{N^{t}}
\left(\begin{array}{cccc} \frac{a_{f, 1}^2}{\sum_{i=1}^n a_{f, i}^2} & 0 & \ldots & 0 \\
0 & \frac{a_{f, 2}^2}{\sum_{i=1}^n a_{f, i}^2} & \ldots & 0 \\
\vdots & \vdots & \ddots & \vdots \\
0 & 0 & \ldots & \frac{a_{f, N}^2}{\sum_{i=1}^n a_{f, i}^2}
\end{array} \right). \]
The error from replacing $E_f$ by $E'_f$ is
\[ \sum_{f} \frac{1}{N^t} \left| 1- \frac{1}{\sum_{i=1}^N a_{f, i}^2} \right| .\]
By applying equation (\ref{eq:Cheb}) with $C=N^{1/6}$, this is of
the order $O(1/N^{1/3})$.
}

The POVM $E_f$ can be implemented as follows.
Let $c_1, \ldots, c_{t-1}$ be arbitrary and let
$f_j(x)=c_{t-1} x^{t-1}+\ldots+c_1 x + j$.
Then, for each $i$ and each $l\in \{0, \ldots, N-1\}$,
there is exactly one $j$ such that $f_j(i)=l$.
This means that, for each $i$,
\[ \sum_j a_{f_j, i}^2= N \sum_{l=1}^{2t} q_l \frac{x^2_l}{N} = 1 \]
and $\sum_j E_{f_j} = \frac{I}{N^{t-1}}$.

Thus, $(c_1, \ldots, c_{t-1})$ is a uniformly random $(t-1)$-tuple
of elements of $\{0, 1, \ldots, N-1\}$.
Let $\ket{\psi}=\sum_{j=0}^{N-1} \alpha_j \ket{j}$ be the state
that is being measured w.r.t. $E_f$.
To measure $c_0$ (after ``measuring'' $c_1, \ldots, c_{t-1}$), we first
create an ancilla state
\[ \sum_{l=1}^{2t} \frac{x_l}{\sqrt{N}}
\sum_{i=N (q_1+q_2+\ldots+q_{l-1}) +1}^{
N (q_1+q_2+\ldots+q_{l})} \ket{i}.\]
We then compute $m=c_{t-1} j^{t-1}+ \ldots + c_1 j$,
perform the transformation $\ket{i}\rightarrow \ket{i-m}$,
uncompute $m$ and measure $c=i-m$.

Conditional on obtaining outcome $c$, this results in the
transformation $U_c\ket{j} = \frac{x_l}{\sqrt{N}}\ket{j}$
where $l$ is such that
\[ q_1+\ldots+q_{l-1}< \frac{f_c (j)}{N} \leq q_1+ \ldots+q_l .\]
By definition of $a_{f, j}$, we have $a_{f_c, j}= \frac{x_l}{\sqrt{N}}$.
Thus, $U_c\ket{j}=a_{f_c, j}\ket{j}$.
The corresponding measurement operator is $E_c=U_c U_c^{\dagger}$.
We have $E_{f_c}=\frac{1}{N^{t-1}} E_c$.
Thus, taking $c_0=c$ results in a correct implementation of
the POVM $E_f$.

\comment{ After $f$ has been measured, we just have to perform a
measurement w.r.t. POVM consisting of
$\frac{1}{N^{2t-1}}\ket{\psi_g}\bra{\psi_g}$ where $\ket{\psi_g}=
\frac{1}{\sqrt{N}} \sum_{l=1}^N e^{2\pi i \frac{g(j)}{N} }\ket{j}$
and $g(x) =d_{2t-1} x^{2t-1} + \ldots + d_1 x + d_0$. } Next, we
show how to measure $E_g$. Let $d_0$, $d_2, \ldots, d_{t-1}$ be
arbitrary but fixed and $g_j(x)=d_0 + j x + d_2 x^2 +\ldots +
d_{2t-1} x^{2t-1}$.

Let $U_{g_0}\ket{j} = e^{2\pi i \frac{g_0(j)}{N}} \ket{j}$.
Then, $U_{g_0}\ket{\psi'_l}=\ket{\psi_{g_l}}$ where
\[ \ket{\psi'_l} = \frac{1}{\sqrt{N}} \sum_{j=0}^{N-1}
e^{-2\pi i \frac{jl}{N}} \ket{j} \]
are just the
vectors of Fourier basis. This has two consequences.
First,
\[ \sum_{l=0}^{N-1} \ket{\psi_{g_l}} \bra{\psi_{g_l}} =
U_{g_0} \left(\sum_{l=0}^{N-1} \ket{\psi'_l}
\bra{\psi'_l} \right) U^{\dagger}_{g_0} =
U_{g_0} I U^{\dagger}_{g_0} = I .\]
Therefore, $(d_0, d_2, \ldots, d_{2t-1})$ is
just a uniformly random vector of $2t-1$ values
from $\{0, \ldots, N-1\}$ which can be generated by
producing the uniform superposition of all
$\ket{d_0, d_2, d_3, \ldots, d_{2t-1}}$ and measuring it.

Second, once $(d_0, d_2, \ldots, d_{2t-1})$ has been produced,
$d_1$ can be obtained by performing $U^{\dagger}_{g_0}$ and
then an orthogonal measurement in the basis
$\ket{\psi'_1}$, $\ldots$, $\ket{\psi'_N}$.

\end{appendix}
\end{document}